\documentclass[a4paper,12pt]{article}
\usepackage{amsmath,epsfig}

\begin{document}
\hspace*{12.4cm}IU-MSTP/65 \\
\hspace*{13cm}hep-lat/0501008 \\
\hspace*{13cm}January, 2005

\begin{center}
{\Large\bf Canonical Approach to Ginsparg-Wilson Fermion}
\end{center}

\vspace*{1cm}
\def\thefootnote{\fnsymbol{footnote}}
\begin{center}{\sc Kosuke Matsui${}^1$, Tomohiro Okamoto${}^1$  
and Takanori Fujiwara${}^2$}
\end{center}
\vspace*{0.2cm}
\begin{center}
{\it ${}^1$ Graduate School of Science and Engineering, Ibaraki University, 
Mito 310-8512, Japan \\
${}^2$ Department of Mathematical Sciences, Ibaraki University, 
Mito 310-8512, Japan}
\end{center}

\vfill
\begin{center}
{\large\sc Abstract}
\end{center}
Based upon the lattice Dirac operator satisfying the Ginsparg-Wilson relation, 
we investigate canonical formulation of massless fermion on the spatial lattice. 
For free fermion system exact chiral symmetry can be implemented without species 
doubling. In the presence of gauge couplings the chiral symmetry is violated. 
We show that the divergence of the axial vector current is related to the chiral 
anomaly in the classical continuum limit.  
\vskip .3cm
\noindent
{\sl PACS:} 11.15.-q, 11.15.Ha, 12.38.Gc

\noindent
\newpage
\pagestyle{plain}
\section{Introduction}
\label{sec:intro}
\setcounter{equation}{0}

The discovery of lattice Dirac operators satisfying the Ginsparg-Wilson 
(GW) relation \cite{gw,has,neub} enables us to implement exact chiral 
symmetry on the lattice \cite{hln,lus} 
without suffering from species doubling \cite{Nielsen-Ninomiya}. 
The overlap Dirac operator \cite{neub}, for 
instance, not only possesses all the desired classical properties such as 
the continuum limit and locality \cite{hjl} but also reproduces correct chiral 
anomaly \cite{chiral_anom,fns}. The lattice Dirac action with the exact chiral symmetry 
may be considered as a correct starting point of the nonperturbative 
studies of gauge theories with massless fermions \cite{lus2,lus3}. 

In lattice gauge theories it is usual to employ euclidean path integral 
formalism. The main concern there is to compute various physical 
quantities nonpertubatively. To investigate formal aspects of the 
theory concerning states and operators it is suitable to 
work in canonical approach. To the author's knowledge, canonical 
treatment of GW fermion has only been investigated by Creutz, 
Horv\'ath and Neuberger \cite{chn}. They found an interesting characteristic 
structure of the spectra of energy and axial charge leading to axial 
anomaly. In this paper we will pursue the validity of their approach 
further and establish the chiral anomaly in the classical continuum limit. 

On the full eucliean lattice, it is possible to give action with 
exact chiral invariance at the classical level \cite{lus}. The fermion 
measure, however, is not chiral invariant but gives raise to nontrivial 
Jacobian \cite{Fujikawa,lus}. It reproduces chiral anomaly in the 
classical continuum limit \cite{chiral_anom,fns}. In the canonical 
approach time is a continuous coordinate and only the spatial 
coordinates are discretized. This causes a problem in constructing 
chirally symmetric action. 

To put this more precisely, we consider a Dirac 
operator\footnote{See Sect. \ref{sec:cwgf} for notation.}
\begin{eqnarray}
  \label{eq:nldo}
  D=i\gamma^0{\cal D}_0+D_{d-1}~, 
\end{eqnarray}
where ${\cal D}_0$ is the time component of the ordinary covariant 
derivative in the continuum and $D_{d-1}$ is the spatial part of 
$D$. It is assumed to satisfy the GW relation. If we choose $D_{d-1}$ 
to be the overlap operator, species doublers can be avoided. 
The full Dirac operator $D$, however, does not satisfy the GW relation 
even in the absence of gauge couplings. Nevertheless it is possible to 
define exact chiral symmetry in the free field case. The chiral 
transformation there is analogous to the one given in Ref. \cite{lus}.
The axial charge of a free GW fermion depends on momentum. 
In the physical momentum region it is almost constant as in the continuum 
theory. The separation of positive and negative axial charges decreases as 
the energy increases. It vanishes at the doubler momenta. In other words 
the states with positive axial charge and those with negative one are 
smoothly connected with one another at the boundaries of the first 
Brillouin zone \cite{chn}. 

In the presence of gauge couplings, the chiral transformation becomes 
field dependent. It is no more a symmetry of the action. This is quite 
different with the conventional euclidean approach. The violation of 
the chiral symmetry is not a bad news. We expect 
it from the beginning since the chiral symmetry must be broken in any 
regulalized theory, otherwise we could not reproduce the chiral anomaly. 
We show that certain gauge fields induce asymmetric flow in the spectrum 
that leads to nonconservation of the axial charge. This is the origin of 
chiral anomaly. 

The interpretation of axial anomaly that external gauge 
fields can generate asymmetric level shifts of the negative energy 
states in the Dirac sea is well-known in the continuum theory \cite{nn}\footnote{There 
are several unpublished works. See references cited in \cite{agp,Manohar}. 
Quantized field theory approach was investigated in 
Ref. \cite{Fujiwara-Ohnuki}.}. Similar analysis has been be carried out also 
for Wilson fermion. In this case the chiral symmetry, however, is broken break 
by the Wilson term even in the absence of gauge couplings. The axial anomaly for 
Wilson fermions was obtained in Ref. \cite{ks} and its physical picture 
was investigated in Ref. \cite{agp}. 

This paper is organized as follows. In the next section we consider 
free GW fermion and introduce exact chiral symmetry. In Sect. 
\ref{sec:cwgf} we examine the effects of couplings with the gauge 
fields on the chiral symmetry and give the axial current divergence. 
In Sect. \ref{sec:cq} we describe the canonical approach 
explicitly for a two-dimensional system and examine the conservation 
of the axial charge for an adiabatically changing external electric 
field. In Sect. \ref{sec:ca} we compute the classical continuum limit 
of the axial current divergence in an arbitrary smooth background and 
show that the chiral anomaly is correctly reproduced. 
Sect. \ref{sec:sd} is devoted to summary and discussion. We collect 
notation in Appendix \ref{sec:not}. The technical detail in carrying 
out the momentum integrations to compute the anomaly coefficients 
is given in Appendix \ref{sec:mi}.

\section{Exact Chiral Symmetry}
\label{sec:odo}

\setcounter{equation}{0}

In this section we consider canonical formulation of a lattice Dirac theory in an 
arbitrary even dimensional space-time with continuous time $t$ and $d-1$ 
discretized lattice coordinates $x$. The signature is assumed to be minkowskian and 
the spatial part is hypercubic regular lattice with lattice spacing 
$a$. The massless fermion action can be generically written as
\begin{eqnarray}
  \label{eq:lda}
  S_F&=&a^{d-1}\sum_x\int dt\overline\psi\Biggl(i\gamma^0
  \frac{\partial}{\partial t}+D_{d-1}\Biggr)\psi~, 
\end{eqnarray}
where $D_{d-1}$ stands for the spatial part of the lattice 
Dirac operator. Unlike euclidean formulation $\psi$ and 
$\overline\psi$ are related by the Dirac conjugate
\begin{eqnarray}
  \label{eq:dc}
  \overline\psi=\psi^\dagger\gamma^0~. 
\end{eqnarray}
The conventions for the metric and $\gamma$-matrices 
are summarized in Appendix \ref{sec:not}. 

To keep the fermion massless while avoiding species 
doubling, we employ the following construction analogous to the overlap 
operator  
\begin{eqnarray}
  \label{eq:odo}
  D_{d-1}=\frac{1}{a}\Biggl(1+\gamma_{d+1}
  \frac{H_{d-1}}{\sqrt{H_{d-1}^2}}\Biggr)~, 
\end{eqnarray}
where $H_{d-1}$ is the hermitian Wilson-Dirac operator given by 
\begin{eqnarray}
  \label{eq:hwdo}
  H_{d-1}=\gamma_{d+1}(aD_W^{(d-1)}-m)~, \qquad
  D_W^{(d-1)}=\sum_{k=1}^{d-1}(i\gamma^k\partial_k^S
  -r\partial_k^A)~.
\end{eqnarray}
The $\partial_k^{S}$ and  $\partial_k^{A}$ are, respectively, symmetric 
and antisymmetric difference operators defined by (\ref{eq:sasdo}) 
and $D_W^{(d-1)}$ is nothing but the Wilson-Dirac operator with 
the difference along the $d$-th (euclidean time) axis omitted. 
This choice is consistent with the cubic symmetry of the underlying 
lattice. The $r$ and $m$ are parameters. They must be so chosen that 
the theory contains no doublers. Here we assume $r>0$. 

The Dirac operator (\ref{eq:odo}) satisfies the following relations 
\begin{eqnarray}
  \label{eq:gwr}
  \gamma_{d+1}D_{d-1}+D_{d-1}\gamma_{d+1}=aD_{d-1}\gamma_{d+1}
  D_{d-1}~, \qquad
  \gamma^0D_{d-1}+D_{d-1}\gamma^0=aD_{d-1}\gamma^0
  D_{d-1}~, 
\end{eqnarray}
where the first is the Ginsparg-Wilson relations as in the case of overlap 
Dirac operator on full euclidean lattices.  The second one is specific to 
the present canonical approach. It can be seen by noting the fact that 
$D_{d-1}$ commutes with $\gamma^0\gamma_{d+1}
=i^{\frac{d}{2}-1}\gamma^1\dots\gamma^{d-1}$. 
It also satisfies the following hermiticity
\begin{eqnarray}
  \label{eq:2kh}
  D_{d-1}^\dagger\gamma^0=\gamma^0D_{d-1}~, \quad
  D_{d-1}^\dagger\gamma_{d+1}=\gamma_{d+1}D_{d-1}~. 
\end{eqnarray}
From the action (\ref{eq:lda}) we define single-particle hamiltonian 
$h$ by
\begin{eqnarray}
  \label{eq:qmh}
  h=-\gamma^0D_{d-1}~.
\end{eqnarray}
The hermiticity of $h$ can be seen from the $\gamma^0$ hermiticity 
(\ref{eq:2kh}).  

In the absence of the interaction with gauge field the action 
(\ref{eq:lda}) is invariant under the lattice chiral transformation
\begin{eqnarray}
  \label{eq:lctr}
  \delta\psi=i\epsilon\gamma_{d+1}\Biggl(1-\frac{a}{2}D_{d-1}\Biggr)
  \psi~, \qquad
  \delta\overline\psi=i\epsilon\overline\psi
  \Biggl(1-\frac{a}{2}D_{d-1}\Biggr)\gamma_{d+1}~,
\end{eqnarray}
where $\epsilon$ is an arbitrary real parameter. These chiral 
transformations are consistent with the Dirac conjugate (\ref{eq:dc}) and the 
hermiticity of $D_{d-1}$. 

The invariance of the action under the chiral symmetry also 
implies that the axial charge operator $q_{d+1}$ defined by 
\begin{eqnarray}
  \label{eq:co}
  q_{d+1}=\gamma_{d+1}\Biggl(1-\frac{a}{2}D_{d-1}\Biggr)~.
\end{eqnarray}
commutes with $h$. Furthermore, $h$ and $q_{d+1}$ satisfy \cite{chn}
\begin{eqnarray}
  \label{eq:cr}
  \frac{a^2}{4}h^2+q_{d+1}^2=1~.
\end{eqnarray}
These can be checked directly by using (\ref{eq:gwr}) 
and (\ref{eq:2kh}). We thus see that $(aE/2,Q_{d+1})$ with $E$ and 
$Q_{d+1}$, respectively, the eigenvalues of $h$ and $q_{d+1}$ can be regarded 
as a point on the unit circle and unlike the continuum theory the 
axial charge depends on the energy of the state. 

To see the doubling problem we go over to momentum representation 
\begin{eqnarray}
  \label{eq:mes}
  \psi(t,x)=u(p)e^{i(px-Et)}~. \qquad
  \Bigl(px=\sum_{k=1}^{d-1}p^kx^k\Bigr)
\end{eqnarray}
To parametrize the eigenvalues of $h$ and $q_{d+1}$ it is convenient to introduce 
$d$-dimensional orthonormal coordinates $n$ by 
\begin{eqnarray}
  \label{eq:ddimc}
  &&n^k=\rho\sin ap^k~, \qquad
  n_d=-\rho\Bigl(r\sum_{k=1}^{d-1}(\cos ap^k-1)+m\Bigr)~,  
  \nonumber \\
  &&\rho^{-1}=\sqrt{\sum_{k=1}^{d-1}\sin^2ap^k+\Biggl(r\sum_{k=1}^{d-1}
  (\cos ap^k-1)+m\Biggr)^2}~.
\end{eqnarray}
These define a mapping from $T^{d-1}$, the 1st  Brillouin zone $-\frac{\pi}{a}<p^k
\leq\frac{\pi}{a}$ with the opposite boundaries identified, to $S^{d-1}$. 
The energy eigenvalues $E$ and the axial charge $Q_{d+1}$ are given by
\begin{eqnarray}
  \label{eq:eev}
  \frac{a}{2}E=\pm\sqrt{\frac{1+n_d}{2}}~, \qquad
  Q_{d+1}=\pm\sqrt{\frac{1-n_d}{2}}~.
\end{eqnarray}
Doublers appear at $p^k=0$ or $\pi/a$ except for  the origin $p^k=0$ if 
$n_d=-1$ is simultaneously satisfied. For $0<m/r<2$ the energy $E$ vanishes 
only at the origin . 

In quantum theory we impose the equal-time anticommutation relations
\begin{eqnarray}
  \label{eq:etacr}
  \{\psi_\alpha(t,x),\psi^\dag_\beta(t,y)\}=\delta_{\alpha\beta}
  \delta_{x,y}/a^{d-1}~, \quad
  \{\psi_\alpha(t,x),\psi_\beta(t,y)\}=
  \{\psi^\dag_\alpha(t,x),\psi^\dag_\beta(t,y)\}=0~,
\end{eqnarray}
where $\alpha$ and $\beta$ are spinor indices. The hamiltonian, the fermion number 
and the axial charge are given by
\begin{eqnarray}
  \label{eq:ham}
  {\cal H}&=&a^{d-1}\sum_x\psi^\dagger h\psi~, \\
  \label{eq:Q}
  {\cal Q}&=&a^{d-1}\sum_x\psi^\dag \psi~,\\
  \label{eq:q5}
  {\cal Q}_{d+1}&=&a^{d-1}\sum_x\psi^\dag q_{d+1}\psi~. 
\end{eqnarray}
By expanding the field operators in terms of plane wave solutions 
with definite energy and axial charge, we can introduce creation 
and annihilation operators. In Sect. \ref{sec:cq} we will explicitly carry out 
this in $(1+1)$-dimensions. 

\section{Coupling with Gauge Fields}
\label{sec:cwgf}
\setcounter{equation}{0}

We now introduce the coupling with gauge field. This can be 
achieved by simply replacing the time-derivative with
the covariant derivative ${\cal D}_0=\partial/\partial t+A_0$ and 
the differences $\partial_k$, $\partial_k^\ast$ with the covariantized operators 
$\nabla_k$ and $\nabla_k^\ast$ defined by (\ref{eq:cdos}). 
The fermion action (\ref{eq:lda}) is then replaced by
\begin{eqnarray}
  \label{eq:ldawgc}
  S_F&=&a^{d-1}\sum_{x}\int dt~\overline\psi(
  i\gamma_0{\cal D}_0+D_{d-1})\psi~.
\end{eqnarray}
This is invariant under the gauge transformation
\begin{eqnarray}
  \label{eq:gtr}
  \psi(t,x)&\rightarrow& \Lambda(t,x)\psi(t,x) \nonumber \\ 
  A_0(t,x)&\rightarrow& \Lambda(t,x)A_0(t,x)\Lambda^\dag(t,x)
  -\partial_t\Lambda(t,x)\Lambda^\dagger(t,x) \\
  U_k(t,x)&\rightarrow&\Lambda(t,x)U_k(t,x)\Lambda^\dagger
  (t,x+a\hat k)~, \nonumber
\end{eqnarray}
where $\hat k$ is the unit vector in the $k$-the direction and $U_k(t,x)$ is 
the link variable associated with the link $(x,x+a\hat k)$. 

If we consider the gauge fields as dynamical, we must also introduce their 
kinetic part to the action. The magnetic part of the field strengths can 
be defined from the standard plaquette variables
\begin{eqnarray}
  \label{eq:pv}
  P_{kl}(t,x)=U_k(t,x)U_l(t,x+a\hat k)U_k^\dag (t,x+a\hat l)U_l^\dag (t,x)~. 
\end{eqnarray}
The electric field is given by 
\begin{eqnarray}
  \label{eq:ef}
  E_{k}(t,x)=-\nabla_kA_0(t,x)+\frac{1}{a}\dot U_k(t,x)U_k^\dag(t,x)~. 
\end{eqnarray}
The Wilson action for the gauge field takes the form 
\begin{eqnarray}
  \label{eq:gfk}
  S_G=-a^{d-1}\sum_x\int dt \frac{1}{g^2}{\rm tr}(E_k)^2
  -a^{d-1}\sum_x\int dt\sum_{k<l}\frac{1}{a^4g^2}{\rm tr}(2-P_{kl}-P_{kl}^\dag)~,
\end{eqnarray}
where $g$ is the coupling constant. 

In general the overlap Dirac operator becomes singular for gauge fields where the 
hermitian Wilson-Dirac operator has a zero-mode. This implies that the fermion 
action (\ref{eq:ldawgc}) is not a well-defined functional over the entire space of 
the link variables. It is well-known in the euclidean path integral formulation 
that the overlap Dirac operator is well-defined for gauge field 
satisfying so-called admissibility condition 
\begin{eqnarray}
  \label{eq:amc}
  ||1-P_{kl}||<\varepsilon~,
\end{eqnarray}
where $||\cdots||$ stands for operator norm and $\varepsilon$ is some positive 
constant depending on the parameters $r$, $m$ and the dimensionality 
$d$ \cite{hjl}. 
By excising nonadmissible configurations the space of lattice gauge fields 
acquires nontrivial topological structure. The admissibility condition 
helps to define the chiral and gauge anomalies precisely in the full euclidean 
lattice theory. 

Practically, nonadmissible gauge fields can be avoided by modifying the gauge 
field action so that they are decoupled from the system \cite{lus2}. Incidentally, in $(1+1)$ 
dimensions there is a parameter region of $r$ and $m$ where the overlap Dirac 
operator $D_1$ can be defined for any gauge field. This case is of great 
interest since the system corresponds to the lattice regularization of 
the massless Schwinger model. In the remainder of this paper we shall 
consider the lattice gauge fields as background external field. 

The chiral transformation (\ref{eq:lctr}), however, is not consistent with the 
coupling of the gauge field. The spatial part of the fermion action (\ref{eq:ldawgc}) 
is invariant by construction. However, the term containing time-derivative violates 
the symmetry. The difficulty comes from the noncommutativity 
of ${\cal D}_0$ and $D_{d-1}$. In fact under the variation (\ref{eq:lctr}) with $\epsilon$ 
being an arbitrary local parameter the action changes as
\begin{eqnarray}
  \label{eq:dS}
  \delta S_F&=&ia^{d-1}\sum_{x}\int dt~\overline\psi
  \Biggl(i\gamma_0\gamma_{d+1}\dot\epsilon\Bigl(1-\frac{a}{2}D_{d-1}\Bigr)
  -\frac{ia}{2}\gamma_0\gamma_{d+1}\epsilon[{\cal D}_0,D_{d-1}] \nonumber \\
  &&\hskip 3cm +[\epsilon,D_{d-1}]\gamma_{d+1}\Bigl(-1+\frac{a}{2}D_{d-1}
  +\frac{a}{2}(i\gamma_0{\cal D}_0+D_{d-1})\Bigr)\Biggr)\psi~.
\end{eqnarray}
If $\epsilon$ is a constant parameter, the first and the last term in the 
integrand are absent. Hence the chiral symmetry is violated if $[{\cal D}_0,D_{d-1}]$ 
is nonvanishing. Incidentally, we can retain the chiral invariance for static 
gauge fields in the temporal gauge $A_0=0$. 

The variation (\ref{eq:dS}) of the action leads to the axial current divergence 
relation
\begin{eqnarray}
  \label{eq:bavcd}
  \frac{\partial}{\partial t}\Bigl\{\overline\psi\gamma_0\gamma_{d+1}
  \Bigl(1-\frac{a}{2}\overrightarrow D_{d-1}\Bigr)\psi\Bigr\}
  -i\overline\psi(\overrightarrow D_{d-1}-\overleftarrow D_{d-1})\gamma_{d+1}
  \Bigl(1-\frac{a}{2}\overrightarrow D_{d-1}\Bigr)\psi
  =-\frac{a}{2}\overline\psi\gamma_0\gamma_{d+1}[{\cal D}_0,\overrightarrow D_{d-1}]\psi~,
  \nonumber \\
\end{eqnarray}
where we have introduced the following notation to simplify the expressions 
\begin{eqnarray}
  \label{eq:oa}
  \overrightarrow D_{d-1}\psi=D_{d-1}\psi(t,x)~, \qquad
  \overline\psi\overleftarrow D_{d-1}=\overline\psi D_{d-1}(t,x)~.
  \end{eqnarray}
The axial charge density appearing in the time derivative of this expression 
coincides with (\ref{eq:q5}). 
The second term on the lhs of (\ref{eq:bavcd}) corresponds to spatial 
divergence of the axial current. In fact the integral of this term over the 
spatial lattice vanishes. This can be seen by noting the relation like  
\begin{eqnarray*}
  a^{d-1}\sum_{x}\overline\psi
  \overleftarrow D_{d-1} \psi=a^{d-1}\sum_{x}\overline\psi
  \overrightarrow D_{d-1} \psi=a^{d-1}\sum_{x}\overline\psi
  D_{d-1} \psi~. 
\end{eqnarray*}

What we have seen is that in the presence of the couplings with an external 
gauge fields the commutativity of $D_{d-1}$ with the time derivative is lost
and consequently the chiral symmetry (\ref{eq:lctr}) is violated. In Sect. 
\ref{sec:ca} we will show that the violation of the chiral symmetry, the 
rhs of (\ref{eq:bavcd}), which is naively of ${\rm O}(a)$, reproduces the 
chiral anomaly in the classical continuum limit.

Since the chiral transformation (\ref{eq:lctr}) is not an exact symmetry, 
we may consider naive chiral transformations 
\begin{eqnarray}
  \label{eq:nctr}
  \delta\psi=i\epsilon\gamma_{d+1}\psi~, \qquad
  \delta\overline\psi=i\epsilon\overline\psi\gamma_{d+1}, 
\end{eqnarray}
which is not a symmetry of the action as well. We close this section with a comment 
on what happens if one uses naive chiral transformations instead of (\ref{eq:lctr}) 
in the evaluation of chiral anomaly. The variation of the action (\ref{eq:ldawgc}) 
under the naive chiral transformation gives the following axial current 
divergence  
\begin{eqnarray}
  \label{eq:dsunctr}
  \frac{\partial}{\partial t}\{\overline\psi\gamma^0\delta_{d+1}\psi\}
  -i\overline\psi(\overrightarrow D_{d-1}-\overleftarrow D_{d-1})\psi
  =-ia\overline\psi\overrightarrow D_{d-1}\gamma_{d+1} D_{d-1}\psi. 
\end{eqnarray}
Compared with (\ref{eq:bavcd}), we arrive at an apparently different form of 
chiral anomaly if we employ (\ref{eq:nctr}) as the chiral transformation. 
These two, however, differ only by a total divergence of some gauge invarinat 
current $k^\mu$ as can be seen from 
\begin{eqnarray}
  \label{eq:tarel}
  -ia\overline\psi\overrightarrow D_{d-1}\gamma_{d+1} D_{d-1}\psi
  =-\frac{a}{2}\overline\psi\gamma_0\gamma_{d+1}[{\cal D}_0,\overrightarrow D_{d-1}]\psi
  +\dot k^0+\partial_l^\ast k^l
\end{eqnarray}
where $k^\mu$ is defined by the relations 
\begin{eqnarray}
  \label{eq:divk}
  k^0=\frac{a}{2}\overline\psi\gamma_0\gamma_{d+1}
  \overrightarrow D_{d-1}\psi, \qquad
  \partial_l^\ast k^l
  =-\frac{ia}{2}\overline\psi(\overrightarrow D_{d-1}-\overleftarrow D_{d-1})\gamma_{d+1}
  \overrightarrow D_{d-1}\psi \quad(l=1,2,3).
\end{eqnarray}
The axial currents corresponding to the transformations (\ref{eq:lctr}) and 
(\ref{eq:nctr}) differ only by a gauge invariant current of O$(a)$ and give 
essentially the same chiral anomaly in the classical continuum limit.

\section{Ginsparg-Wilson Fermion in $(1+1)$-dimensions}
\label{sec:cq}

\setcounter{equation}{0}

In this section we apply the formalism developed in the preceding sections to the 
case of $(1+1)$-dimensions. We consider a finite periodic lattice of size $L=a(2N+1)$, 
where we have assumed that the number of sites is odd. This is only a 
technical assumption to make the arguments simple.

In momentum space the eigenspinors satisfy the Dirac equation 
\begin{eqnarray}
  \label{eq:deims}
  \gamma^0Eu(p)+D_1(p)u(p)=0~, 
  \qquad
  D_1(p)=\frac{1}{a}(\gamma^1 n_1+n_2+1)~,
\end{eqnarray}
where $D_1(p)$ is the overlap operator in momentum space. In $(1+1)$-dimensions 
the eigenspinors can be more conveniently parametrized by a pseudo-momentum $\tilde p$ 
defined by
\begin{eqnarray}
  \label{eq:pm}
  &&\sin a\tilde p=-n_1=\frac{\sin ap}{\sqrt{\sin^2ap+(r(\cos ap-1)+m)^2}}~, 
  \nonumber \\
  &&\cos a\tilde p=-n_2=\frac{r(\cos ap-1)+m}{\sqrt{\sin^2ap+(r(\cos ap-1)+m)^2}}~.
\end{eqnarray}
For $0<m/r<2$ these define one-to-one mapping from $-\pi/a<p\leq\pi/a$ to 
$-\pi/a<\tilde p\leq\pi/a$. We consider $\tilde p$ as continuous function of $p$ 
beyond the first Brillouin zone. In particular we have 
\begin{eqnarray}
  \label{eq:p2ptc}
  \widetilde{p+\frac{2\pi}{a}}=\tilde p+\frac{2\pi}{a}~. 
\end{eqnarray}
One easily see from (\ref{eq:ddimc}) that the choice $r=m=1$ leads to $\tilde p=p$.  
In $(1+1)$-dimensions this parameter choice is special in the sense that the overlap 
operator $D_1$ becomes identically equal to the Wilson operator. This can be verified 
directly by establishing the relation $H_1^2=1$. Remarkably, it holds true even in 
the presence of gauge couplings. 

The eigenspinor for $E\ne0$ is automatically eigenspinor for the chiral operator $q_3$ 
defined by (\ref{eq:co}). We thus find plane wave solutions 
for $\displaystyle{E=E_p=\frac{2}{a}\sin{\frac{a\tilde p}{2}}}$ and $\displaystyle{Q_3
=\chi_p=\cos\frac{a\tilde p}{2}}$ 
\begin{eqnarray}
  \label{eq:uR}
  U_{Rp}(t,x)=\frac{1}{\sqrt{L}}u_R(p)e^{i(px-E_pt)} ~, \qquad
  u_R(p)=\left(\begin{matrix}\displaystyle{\cos\frac{a\tilde p}{4}} \\ 
      \displaystyle{-\sin\frac{a\tilde p}{4}}
    \end{matrix}\right)
\end{eqnarray}
and for $\displaystyle{E=-E_p=-\frac{2}{a}\sin{\frac{a\tilde p}{2}}}$ and $\displaystyle{
Q_3=-\chi_p=-\cos\frac{a\tilde p}{2}}$ 
\begin{eqnarray}
  \label{eq:uL}
    U_{Lp}(t,x)=\frac{1}{\sqrt{L}}u_L(p)e^{i(px+E_pt)} ~, \qquad
    u_L(p)=\left(\begin{matrix}\displaystyle{\sin\frac{a\tilde p}{4}} \\ 
        \displaystyle{\cos\frac{a\tilde p}{4}}
    \end{matrix}\right)~.
\end{eqnarray}
The dispersion relations $E=\pm E_p$ are easily recognized as the lattice 
analog of $E=\pm p$ known for the continuum chiral theory. 
Unlike the fermion number the conserved axial charge depends on the momentum. 
For the physical region $|p|<\hskip -.1cm<\pi/a$ the axial charge is  
$Q_3=\pm\chi_p\approx\pm1$. It approaches to zero at the boundaries of the 
1st Brillouin zone $|p|\approx \pi/a$. 

\begin{figure}[t]
  \centering
  \epsfig{file=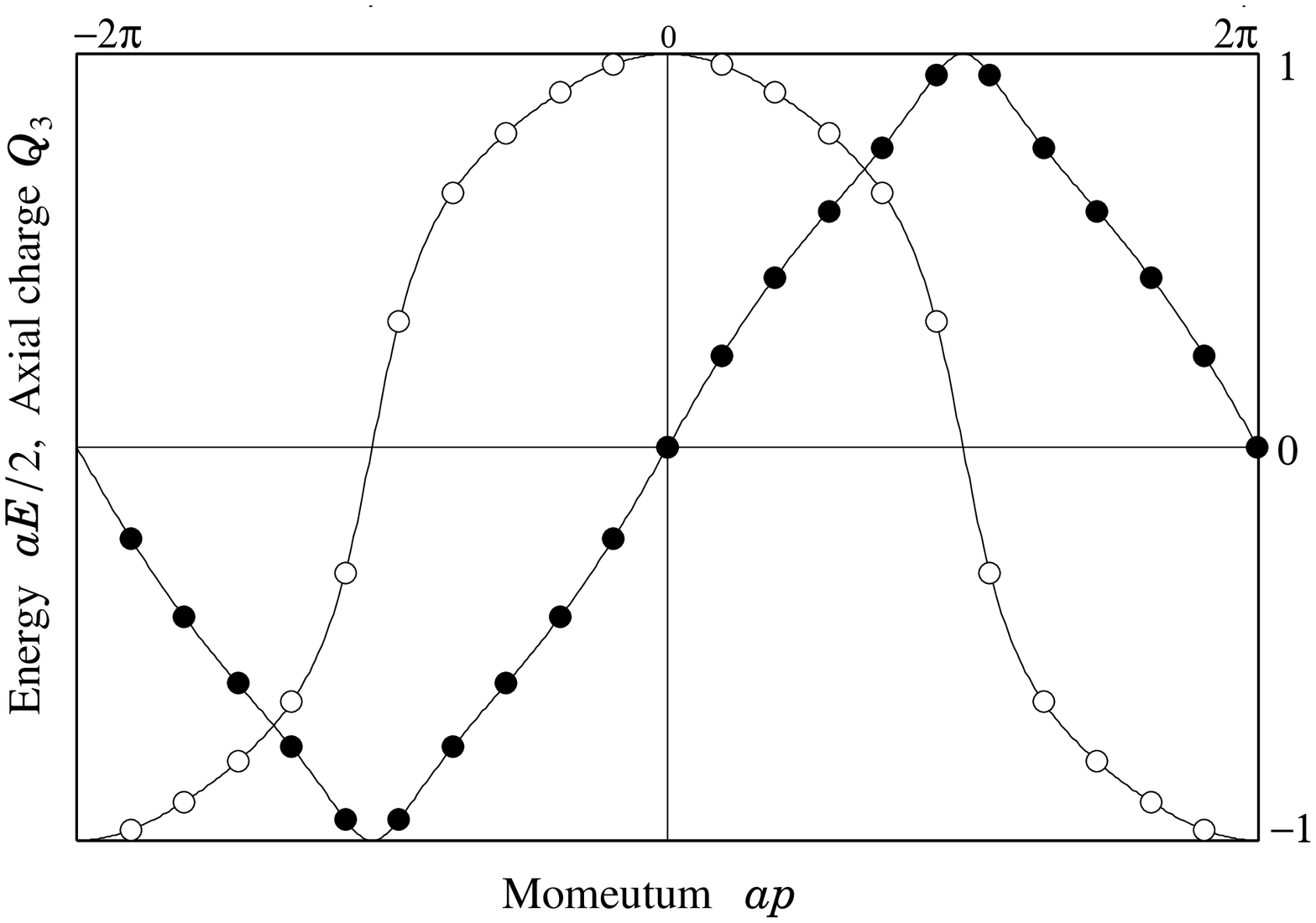,clip=,height=7cm, width=8.5cm,angle=0}
  \epsfig{file=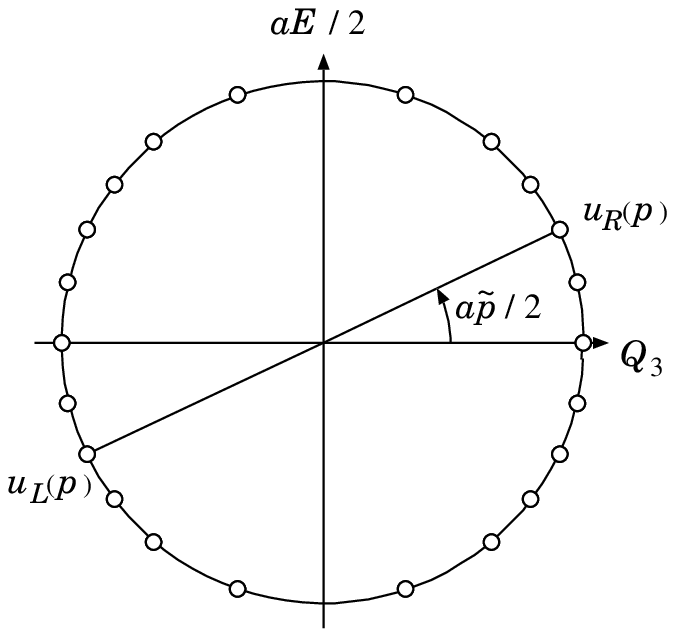,clip=,height=7cm,angle=0}
  \caption{The energy ($\bullet$) and the axial charge ($\circ$) are 
    plotted as functions of momentum for $r=0.8$ and $m=1.2$ with 
    $N=5$ (left). They are conveniently described as points on a 
    unit circle (right) \cite{chn}. Diametrically opposite points 
    correspond to a pair of eigenspinor $u_R(p)$ and $u_L(p)$.}
  \label{fig:1}
\end{figure}

The momentum $p$ is usually restricted to lie in the 1st Brillouin zone. 
In the present chiral theory the period of the spectra of energy and axial 
charge is not $2\pi/a$ but is $4\pi/a$. Furthermore, the wave functions 
(\ref{eq:uR}) and (\ref{eq:uL}) up to an overall sign are transformed into 
each other by the translation $p\rightarrow p+2\pi/a$. As will be clear 
shortly, it turns out to be convenient to double the Brillouin zone and 
use a single wave function defined by 
\begin{eqnarray}
  \label{eq:u}
  U_p(t,x)=\frac{1}{\sqrt{L}}u(p)e^{i(px-E_pt)}~, \qquad
  u(p)=\left(\begin{matrix}\displaystyle{\cos\frac{a\tilde p}{4}} \\ 
      \displaystyle{-\sin\frac{a\tilde p}{4}}
    \end{matrix}\right)~.
\end{eqnarray}
By noting the relation (\ref{eq:p2ptc}) we see that up to 
an overall sign $U_p$ is periodic in $p$ with a period $4\pi/a$ and 
$U_p=U_{Rp}$ for $-\pi<ap<\pi$  and $U_p=U_{Lp-2\pi/a}$ for 
$\pi<ap\leq3\pi$. The relation among the momentum, 
the energy and the axial charge are shown in Fig. \ref{fig:1}. 

The free Dirac field can be expanded in terms of $U_p$ as 
\begin{eqnarray}
  \label{eq:fdfexp}
  \psi(t,x)=\sum_{0\leq ap<4\pi}a_pU_p(t,x)~,
\end{eqnarray}
where $a_p$ is a creation or an annihilation operators satisfying 
the anti-commutation relations
\begin{eqnarray}
  \label{eq:caacrl}
  \{a_p,a^\dagger_{p'}\}=\delta_{p,p'}~, \quad
  \{a_p,a_{p'}\}=\{a^\dag_p,a^\dagger_{p'}\}=0~. \qquad
  (0\leq p,~p'<4\pi/a)
\end{eqnarray}
We see from 
(\ref{eq:u})  that $a_p^\dagger$ creates a particle of 
energy $E_p$ for $0\leq p\leq 2\pi/a$ and  $a_p$ create an anti-particles of 
energy $|E_p|=-E_p$ for $2\pi/a<p<4\pi/a$, where we have assigned the 
zero-energy states as particles for simplicity. 

The hamiltonian $H$, the fermion number ${\cal Q}$ and the axial charge ${\cal Q}_3$ 
are given by 
\begin{eqnarray}
  \label{eq:hamop}
  &&H=a\sum_x\psi^\dagger h\psi=\sum_{0\leq ap<4\pi}
  E_pa_p^\dagger a_p~, \\
  \label{eq:fn}
  &&{\cal Q}=a\sum_x\psi^\dagger\psi=
  \sum_{0\leq ap<4\pi}a_p^\dagger a_p~, \\
  \label{eq:ac}
  &&{\cal Q}_3=a\sum_x\psi^\dagger q_3\psi=
  \sum_{0\leq ap<4\pi}\chi_pa_p^\dagger a_p~.  
\end{eqnarray}
These are not normal ordered with respect to the creation and annihilation 
operators and the energy and the fermion number of the Dirac vacuum are 
nonvanishing. However, the axial charge of the Dirac vacuum vanishes because 
of the cancellation of the axial charge between the occupied negative 
energy states.  

We now introduce $U(1)$ lattice gauge field $(A_0(t,x),U_1(t,x))$ in 
$(1+1)$-dimensions and investigate the behavior of the axial charge. 
In the remainder of this section we assume that the gauge potential is real. The electric 
field defined by (\ref{eq:ef}) is given by 
\begin{eqnarray}
  \label{eq:uef}
  {\cal E}(t,x)=-\partial_1 A_0(t,x)-\frac{i}{a}\partial_t U_1(t,x)
  U_1(t,x)^\dag~,
\end{eqnarray}
which is invariant under the gauge transformation (\ref{eq:gtr}).  In this section 
we are only interested in a spatially uniform electric field generated by 
the gauge field
\begin{eqnarray}
  \label{eq:gpuef}
  A_0(t,x)=0~, \qquad U_1(t,x)=e^{-iaA(t)}~.
\end{eqnarray}
Then the electric field is given by 
\begin{eqnarray*}
  {\cal E}(t)=-\dot A(t)~. 
\end{eqnarray*}
Since the gauge field is translation invariant in the spatial direction, we 
can solve the Dirac equation in momentum space. In the presence 
of the background gauge field the free hamiltonian $-\gamma^0D_1(p)$ 
is modified to  $-\gamma^0D_1(p-A(t))$. Then the Dirac 
equation in momentum space is simply given by 
\begin{eqnarray}
  \label{eq:deqims}
  i\frac{\partial\varphi(t,p)}{\partial t}
  &=&-\gamma^0D_1(p-A(t))\varphi(t,p)~,
\end{eqnarray}
At this point, we further assume that the $A(t)$ changes adiabatically 
from $A(t_i)=0$ to 
\begin{eqnarray}
  \label{eq:q}
  A(t_f)=q=\frac{2\pi n_q}{a(2N+1)}~,
\end{eqnarray}
where $n_q$ is an integer 
satisfying $|n_q|\leq N$. We take $t_f-t_i$ sufficiently large so that the adiabatic 
change of $A(t)$ is possible. In this situation the transitions 
between $u_R$ and $u_L$ are suppressed and a state that is initially in an 
eigenstate of the hamiltonian keeps staying in the corresponding eigenstate of the 
hamiltonian at an arbitrary time. Thus the solution to the Dirac equation 
(\ref{eq:deqims}) satisfying $\varphi(t_i,p)=u(p)$ can be solved in terms of the 
eigenspinor of $-\gamma^0D_1(p-A(t))$ as 
\begin{eqnarray}
  \label{eq:svp}
  &&\varphi(t,p)=u(p-A(t))\exp\Biggl[-i\int_{t_i}^t
  E_{p-A(\tau)}d\tau\Biggr]~, 
\end{eqnarray}
where $u(p)$ is given by (\ref{eq:u}) and $E_p=\displaystyle{\frac{2}{a}
\sin\frac{a\tilde p}{2}}$. Under the influence of the background electric field 
the free Dirac field (\ref{eq:fdfexp}) evolves to   
\begin{eqnarray}
  \label{eq:lh}
  \psi(t,x)=\frac{1}{\sqrt{L}}\sum_{0\le ap<4\pi}a_pu(p-A(t))
  \exp\Biggl[i\Biggl(px-\int_{t_i}^tE_{p-A(\tau)}
  d\tau+\delta_p\Biggr)\Biggr]~.
\end{eqnarray}
The phase $\delta_p$ is chosen so that $\psi(t,x)$
reduces to the free field (\ref{eq:fdfexp}) for $t\leq t_i$. 

For $t\geq t_f$ the Dirac field approaches to 
\begin{eqnarray}
  \label{eq:af}
  \psi(t,x)=\frac{1}{\sqrt{L}}e^{iqx}\sum_{0\leq ap<4\pi}a_pu(p-q)
  e^{i((p-q)x-E_{p-q}t+\delta'_p)}~, 
\end{eqnarray}
where $\delta'_p$  is some irrelevant phase. 

The overall factor $e^{iqx}$ is periodic on the lattice if the condition 
(\ref{eq:q}) is satisfied. It can be removed by carrying out a time-independent 
gauge transformation 
\begin{eqnarray}
  \label{eq:sgtr}
  \psi(t,x)\rightarrow e^{-iqx}\psi(t,x)~. 
\end{eqnarray} 
This transforms $\psi(t,x)$ $(t>t_f)$ into a free field. As an external gauge field, 
it is not necessary to impose the condition (\ref{eq:q}) and any $q=A(t_f)$ is allowed. 
In general $e^{iqx}$, however, is not periodic on the lattice and cannot 
be regarded as a gauge transformaiton on the lattice. Only those satisfying 
(\ref{eq:q}) can be eliminated by the gauge transformaiton (\ref{eq:sgtr}). 

What we found is that the state with initial momentum $p$ is carried 
over to the state with momentum $p-q$ due to the electric field exerted on the 
system. It is well-known in continuum theory that this kind of spectral flow results 
in the violation of the conservation of the axial charge. To see this we compute 
the vacuum expectation value of the axial charge at $t=t_f$. It is given by 
\begin{eqnarray}
  \label{eq:vevq3}
  \langle0|{\cal Q}_3(t_f)|0\rangle&=&\sum_{0\leq ap<4\pi}
  \cos\frac{a\widetilde{(p-q)}}{2}\langle0|a^\dag_pa_p|0\rangle \nonumber \\
  &=&\sum_{0< ap<\pi}\Biggl(
  \cos\frac{a\widetilde{(p+q)}}{2}
  -\cos\frac{a\widetilde{(p-q)}}{2}\Biggr) \nonumber \\   
  &=&\sum_{\pi-2aq< ap<\pi}
  \cos\frac{a\widetilde{(p+q)}}{2}
  -\sum_{0<ap\leq2aq}\cos\frac{a\widetilde{(p-q)}}{2}~. 
\end{eqnarray}
where $|0\rangle$ is the Dirac vacuum satisfying $a_p|0\rangle=0$ for $0\leq
ap\leq2\pi$ and use has been made of (\ref{eq:p2ptc}). In the last equality 
we have assumed $q>0$. The contributions from the 
modes lying near the boundary of the 
Brillouin zone ($\pi-2aq< ap<\pi$) are canceled out and only the modes around 
the origin ($0<ap\leq2aq$) contribute to the violation of the axial charge 
conservation.  We thus find
\begin{eqnarray}
  \label{eq:r0q3r0}
  \langle0|{\cal Q}_3(t_f)|0\rangle&=& -\sum_{-aq< ap\leq aq}
  \cos\frac{a\widetilde p}{2}~.
\end{eqnarray}
The $q<0$ case can be analyzed in a similar way. 

To relate (\ref{eq:r0q3r0}) with the axial anomaly we consider the classical continuum limit 
$a\rightarrow0$ with $L=a(2N+1)$ kept fixed.  For the modes contributing 
to the nonvanishing axial charge we have $\displaystyle{\cos\frac{a\tilde p}{2}\rightarrow 1}$ 
as $a\rightarrow0$. We thus find 
\begin{eqnarray*}
  \langle0|{\cal Q}_3(t_f)|0\rangle\mathop{=}_{a\rightarrow0}-2n_q
  =-\frac{1}{\pi}a(2N+1)q
  =\frac{1}{\pi}a\sum_x\int_{t_i}^{t_f}{\cal E}(t)dt~, \nonumber \\
\end{eqnarray*}
where use has been made of ${\cal E}(t)=-\dot A(t)$. Since 
$\langle0|{\cal Q}_3(t_i)|0\rangle=0$, we finally obtain 
\begin{eqnarray}
  \label{eq:car}
  \int_{t_i}^{t_f}dt\langle0|\dot{\cal Q}_3(t)|0\rangle
  \mathop{=}_{a\rightarrow0}2\frac{1}{2\pi}\int_0^Ldx\int_{t_i}^{t_f}dt{\cal E}(t)~. 
\end{eqnarray}
This coincides with the integrated axial anomaly relation. 

The nonconservation of the axial charge implies the violation of the chiral symmetry. 
In the next section we investigate the chiral symmetry in the presence of an arbitrary 
background gauge field and see how the chiral anomaly is reproduced in the classical 
continuum limit.

\section{Chiral Anomaly}
\label{sec:ca}
\setcounter{equation}{0}

In the presence of a background gauge field the chiral symmetry is violated. 
We will compute the violation of the axial charge by examining the classical 
continuum limit of the vacuum expectation value of the operator appearing in 
the rhs of (\ref{eq:bavcd}). It is a local function of the gauge field 
and is denoted by $2{\cal A}(t,x)$ 
\begin{eqnarray*}
  {\cal A}(t,x)&=&-\frac{a}{4}
  \langle\overline\psi(t,x)\gamma_0\gamma_{d+1}[{\cal D}_0,D_{d-1}]\psi(t,x)\rangle 
  \nonumber \\
  &=&-\frac{a}{4}\lim_{t'\rightarrow t}\lim_{x'\rightarrow x}
  {\rm tr}\gamma_{d+1}\gamma^0
  [{\cal D}_0,D_{d-1}]\langle {\rm T}\psi(t,x)\overline\psi(t',x')\rangle~, 
\end{eqnarray*}
where ${\rm tr}$ stands for trace over spin and internal indices and ${\rm T}$ denotes 
time-ordering. The two-point function $\langle {\rm T}\psi(t,x)\overline
\psi(t',x')\rangle$ 
satisfies 
\begin{eqnarray}
  \label{eq:2pfeq}
  (i\gamma^0{\cal D}_0+D_{d-1})\langle {\rm T}\psi(t,x)\overline\psi(t',x')
  \rangle=i\delta(t-t')\delta_{x,x'}/a^{d-1}~.
\end{eqnarray}
Hence ${\cal A}$ can be written as
\begin{eqnarray}
  \label{eq:ca}
  {\cal A}(t,x)
  &=&-\frac{i}{4a^{d-2}}\lim_{t'\rightarrow t}\lim_{x'\rightarrow x}
  {\rm tr}\gamma_{d+1}\gamma^0
  [{\cal D}_0,D_{d-1}]\frac{1}{i\gamma^0{\cal D}_0+D_{d-1}+i\epsilon}
  \delta(t-t')\delta_{x,x'}~,   
\end{eqnarray}
where the $i\epsilon$ prescription is made explicit. For simplicity 
we assume that the size of the lattice is infinite. The rhs of this 
expression can be simplified by making use of the relation
\begin{eqnarray}
  \label{eq:dfft}
  \delta(t-t')\delta_{x,x'}=\int_{-\infty}^{+\infty}\frac{dE}{2\pi}
  e^{-iE(t-t')}
  \int_{-\pi}^\pi\frac{d^{d-1}p}{(2\pi)^{d-1}}e^{ip(x-x')/a}~. 
\end{eqnarray}
After a bit of algebra, we find  
\begin{eqnarray}
  \label{eq:cangf}
  {\cal A}(t,x)
  &=&-\frac{i}{4a^{d-2}}\int_{-\infty}^{+\infty}\frac{dE}{2\pi}
  \int_{-\pi}^\pi\frac{d^{d-1}p}{(2\pi)^{d-1}}
  {\rm tr}\;\gamma_{d+1}\gamma^0
  [{\cal D}_0,\tilde D_{d-1}]
  \frac{1}{i\gamma^0{\cal D}_0+\gamma^0E+\tilde D_{d-1}+i\epsilon} 
  \nonumber\\
  &=&-\frac{i}{4a^{d-1}}\int_{-\infty}^{+\infty}\frac{dE}{2\pi}
  \int_{-\pi}^\pi\frac{d^{d-1}p}{(2\pi)^{d-1}}
  {\rm tr}\;\gamma_{d+1}\gamma^0
  \Biggl[{\cal D}_0,\gamma_{d+1}\frac{\tilde H_{d-1}}{\sqrt{\tilde H_{d-1}^2}}\Biggr]
  \nonumber \\
  &&\hskip 1cm \times 
  \frac{1}{\gamma^0E+1+\gamma_{d+1}\tilde H_{d-1}/\sqrt{\tilde H_{d-1}^2}
    +ia\gamma^0{\cal D}_0+i\epsilon}~,
\end{eqnarray}
where we have introduced the notation $\tilde {\cal O}=e^{-ipx/a}{\cal O}
e^{ipx/a}$ for any operator ${\cal O}$. In the last equality 
we have used a rescaling trick $aE\rightarrow E$.

To carry out a systematic expansion by the lattice constant $a$ we further 
assume that $A_0(t,x)$ approaches to a smooth function as $a\rightarrow0$ 
and the link variable are given by a smooth gauge potential $A_k(x)$ as 
\begin{eqnarray}
  \label{eq:lv}
  U_k(t,x)={\rm P}\exp\Biggl[a\int_0^1 dsA_k(t,x+a(1-s)\hat k)\Biggr]~. 
\end{eqnarray}
Then the expansions of $\tilde\nabla_k^S$ and $\tilde\nabla_k^A$ are 
given by  
\begin{eqnarray}
  \label{eq:tn}
  a\tilde\nabla_k^S=i\sin p^k+a\cos p^k{\cal D}_k+{\rm O}(a^2)~, \\
  a\tilde\nabla_k^A=\cos p^k-1+ia\sin p^k{\cal D}_k+{\rm O}(a^2)~,
\end{eqnarray}
where ${\cal D}_k=\partial_k+A_k$ is the covariant derivative in the continuum 
theory. For later convenience we use $n_\mu$, 
$m_\mu$ ($\mu=1,\cdots,d$) defined by 
\begin{eqnarray}
  \label{eq:nm}
  && n_k=-\rho\sin p^k~, \quad n_d=-\rho\Bigl(r\sum_k(\cos p^k-1)+m\Bigr)~,
  \nonumber \\
  &&m_k=\rho\cos p^k~, \qquad
  m_d=\rho~, \\
  &&\Biggl(\rho^{-1}=\sqrt{\sum_k\sin^2 p^k
  +\Bigl(r\sum_k(\cos p^k-1)+m\Bigr)^2}~, \quad k=1,\cdots,d-1\Biggr)
\nonumber
\end{eqnarray}
In terms of $n_\mu$ and $m_\mu$ the hermitian Wilson-Dirac 
operator $\tilde H_{d-1}$ and its square can be expanded as 
\begin{eqnarray}
  \label{eq:ehwdop}
  \rho\gamma_{d+1}\tilde H_{d-1}
  =\gamma\cdot n+n_d+V_1~, \qquad
  \rho^2\tilde H_{d-1}^2
  =1+V_2+V_3~,
\end{eqnarray}
where $\displaystyle{\gamma\cdot n=\sum_{k=1}^{d-1}
\gamma^kn_k}$ and $V$'s are given by
\begin{eqnarray}
  \label{eq:er}
  V_1&=&\rho\Bigl(a\sum_ki\gamma^k\tilde\nabla_k^S-r
  a\sum_k\tilde\nabla_k^A-m\Bigr)-\gamma\cdot n-n_d \nonumber \\
  &=&ia\sum_k(\gamma^km_k+rn_k){\cal D}_k+{\rm O}(a^2)~, \nonumber \\
  V_2&=&-a^2\rho^2\sum_k(\tilde\nabla_k^S)^2+\rho^2\Bigl(ar\sum_k
  \tilde\nabla_k^A+m\Bigr)^2-1 \nonumber\\
  &=&2ia\sum_k(m_k+rn_d)n_k{\cal D}_k+{\rm O}(a^2)~, \\
  V_3&=&ira^2\rho^2\sum_{k,l}\gamma^k[\tilde\nabla_k^S,\tilde\nabla_l^A]
  +\frac{1}{2}a^2\rho^2
  \sum_{k,l}\gamma^k\gamma^l[\tilde\nabla_k^S,\tilde\nabla_l^S] \nonumber\\
  &=&a^2\sum_{k,l}\Bigl(\gamma^krm_kn_l+\frac{1}{2}
  \gamma^k\gamma^lm_km_l\Bigr)F_{kl}+{\rm O}(a^3)~.\nonumber 
\end{eqnarray}
In the last line use has been made of the field strength 
$F_{kl}=[{\cal D}_k,{\cal D}_l]$. The reader might think that 
we should expand $V_2$ to order $a^2$. In the actual 
computation the order $a^2$ term in $V_2$ does not 
contribute to ${\cal A}$ at least in two and four dimensions. 

From (\ref{eq:ehwdop}) we obtain 
\begin{eqnarray}
  \label{eq:hoh}
  \gamma_{d+1}\tilde H_{d-1}/\sqrt{\tilde H_{d-1}^2}&=&
  (\gamma\cdot n+n_d+V_1)/\sqrt{1+V_2+V_3}~.
\end{eqnarray}
If we define $V$, which is of order $a$, by 
\begin{eqnarray}
  \label{eq:d}
  V=ia\gamma^0{\cal D}_0
  +\gamma_{d+1}\tilde H_{d-1}/\sqrt{\tilde H_{d-1}^2}
  -\gamma\cdot n-n_d~, 
\end{eqnarray}
we can expand (\ref{eq:cangf}) as 
\begin{eqnarray}
  \label{eq:aD}
  {\cal A}(t,x)&=&\sum_{k=0}^{d-2}{\cal A}^{(k)}+{\rm O}(a)~, 
\end{eqnarray}
where ${\cal A}^{(k)}$ is given by 
\begin{eqnarray}
  \label{eq:ak}
  {\cal A}^{(k)}(t,x)=\lim_{a\rightarrow0}
  (-1)^{k+1}\frac{i}{4a^{d-1}}
  \int_{-\infty}^{+\infty}\frac{dE}{2\pi}
  \int_{-\pi}^\pi\frac{d^{d-1}p}{(2\pi)^{d-1}}
  {\rm tr}\;\gamma_{d+1}\gamma^0
  [{\cal D}_0,V]
  (SV)^kS~.
\end{eqnarray}
The $S$ is the free propagator 
\begin{eqnarray}
  \label{eq:S}
  S=\frac{1}{\gamma^0E+\gamma\cdot n+1+n_d+i\epsilon}~.
\end{eqnarray}
It is now straightforward to compute ${\cal A}^{(k)}$. 

\subsection{$(1+1)$-dimensions}
\label{sec:td}

The computation of (\ref{eq:cangf}) is rather simple 
in two dimensions. From the general argument given above 
only the lowest order term ${\cal A}^{(0)}$ survives in the 
limit $a\rightarrow0$. We thus obtain 
\begin{eqnarray}
  \label{eq:2da}
  {\cal A}(t,x)
  &=&-\frac{i}{4a}\int_{-\infty}^{+\infty}\frac{dE}{2\pi}
  \int_{-\pi}^\pi\frac{dp}{2\pi}
  {\rm tr}\;\gamma_{3}\gamma^0
  \Biggl[{\cal D}_0, V_1-\frac{1}{2}(\gamma\cdot n+n_2)
  V_2\Biggr]
  \nonumber \\
  &&\hskip 1cm \times 
  \frac{1}{\gamma^0E+\gamma\cdot n+1+n_2+i\epsilon}+{\rm O}(a) 
  \nonumber \\
  &=&\frac{i}{4\pi}c_2\epsilon^{\mu\nu}{\rm tr}\;F_{\mu\nu}+{\rm O}(a)~,
\end{eqnarray}
where $c_2$ is a numerical coefficient given by 
\begin{eqnarray}
  \label{eq:2dic2}
  c_2=-\frac{1}{4}\int_{-\pi}^\pi dp
  \frac{m_1(1+n_2)}{\sqrt{2(1+n_2)}}\Bigl(n_2-r\frac{n_1^2}{m_1}
    \Bigr)~. 
\end{eqnarray}
This integral can be evaluated by a change of variable 
$z=\rho\sin p$ as explained in Appendix \ref{sec:mi}. 
We find from (\ref{eq:2di}) $c_2=1$ for $0<m<2r$ and $c_2=0$ 
for $m<0$, $m>2r$. This implies that the chiral symmetry 
breaking (\ref{eq:2da}) approaches to the chiral anomaly of 
the continuum theory for $0<m<2r$. Note also that the 
result is consistent with (\ref{eq:car}).

\subsection{$(1+3)$-dimensions}
\label{sec:4d}

In four dimensions we must compute ${\cal A}^{(k)}$ for 
$k=0,1,2$. Inserting (\ref{eq:d}) into the rhs of 
(\ref{eq:ak}) and carrying out the trace over the 
$\gamma$-matrices and the energy integral, we obtain 
\begin{eqnarray}
  \label{eq:A0}
  {\cal A}^{(0)}&=&\frac{1}{8}
  \int_{-\pi}^\pi\frac{d^3p}{(2\pi)^3}\frac{n_4+1}{\sqrt{2(1+n_4)}}
  \prod_pm_p\Biggl(n_4-r\sum_p\frac{n_p^2}{m_p}\Biggr)
  \sum_{j,k,l}\epsilon^{jkl}{\rm tr}[{\cal D}_0,F_{jk}{\cal D}_l]~,\nonumber \\
  {\cal A}^{(1)}&=&-\frac{1}{48}
  \int_{-\pi}^\pi\frac{d^3p}{(2\pi)^3}
  \frac{5+n_4}{\sqrt{2(1+n_4)}}\prod_pm_p
  \Biggl(n_4-r\sum_p\frac{n_p^2}{m_p}\Biggr)
  \sum_{j,k,l}\epsilon^{jkl}{\rm tr}[{\cal D}_0,F_{jk}]
  {\cal D}_l \nonumber \\
  &&+\frac{1}{48}\int_{-\pi}^\pi\frac{d^3p}{(2\pi)^3}
  \frac{1-n_4}{\sqrt{2(1+n_4)}}\prod_pm_p
  \Biggl(n_4-r\sum_p\frac{n_p^2}{m_p}\Biggr)
  \sum_{j,k,l}\epsilon^{jkl}{\rm tr}F_{0j}F_{kl}~, \\
  {\cal A}^{(2)}&=&\frac{1}{16}
  \int_{-\pi}^\pi\frac{d^3p}{(2\pi)^3}
  \frac{1}{\sqrt{2(1+n_4)}}\prod_pm_p\Biggl(n_4-r\sum_p\frac{n_p^2}{m_p}
  \Biggr)\sum_{j,k,l}\epsilon^{jkl}{\rm tr}F_{0j}F_{kl}~. \nonumber
\end{eqnarray}
We do not reproduce the computation here. Instead, we 
mention some properties useful in getting some feeling about 
the results. Firstly, all the contributions 
which are potentially diverging in the limit $a\rightarrow0$ vanish
by virtue of the $\gamma_5$ in the trace. 
Similarly, the terms of order $a^2$ in $V_1$, $V_2$ and
$ia\gamma^0{\cal D}_0$ in $V$, which should be considered in 
the systematic expansion of $V$ in the lattice constant, can be 
ignored. Then we have only to consider the terms explicitly given 
in (\ref{eq:er}). Secondly, the trace over the spinor indices 
followed by the energy integral yields momentum integrals of 
functions of $m_k$, $n_k$ and $n_4$. Typically they take the forms
\begin{eqnarray}
  \label{eq:mi}
 && \int_{-\pi}^\pi d^3p\sum_{j,k,l}\epsilon^{jkl}m_jm_km_l
  {\cal O}_{jkl}(n_4)=\int_{-\pi}^\pi d^3p\Biggl(\prod_pm_p\Biggr)
  \sum_{j,k,l}\epsilon^{jkl}
  {\cal O}_{jkl}(n_4)~, \nonumber \\
  && \int_{-\pi}^\pi d^3p\sum_{j,k,l}\epsilon^{jkl}m_jm_km_ln_l^2
  {\cal O}'_{jkl}(n_4)=\int_{-\pi}^\pi d^3p\Biggl(\prod_pm_p\Biggr)
  \frac{1-n_4^2}{3}\sum_{j,k,l}\epsilon^{jkl}
  {\cal O}'_{jkl}(n_4)~, \\
  && \int_{-\pi}^\pi d^3p\sum_{j,k,l,p}\epsilon^{jkp}m_jm_kn_pn_j
  {\cal O}''_{jkl}(n_4)=\frac{1}{3}\int_{-\pi}^\pi d^3p
  \Biggl(\prod_pm_p\Biggr)\Biggl(\sum_p
  \frac{n_p^2}{m_p}\Biggr)\sum_{j,k,l}
  \epsilon^{jkl}{\cal O}''_{jkl}(n_4)~. \nonumber 
\end{eqnarray}
In moving from the lhs to rhs of these expressions we have 
used the fact that $n_4$ is symmetric in the momentum variables. 

What remains to show is the evaluation of the momentum integrals.
This is done in Appendix \ref{sec:mi}. Using (\ref{eq:I12}), we find 
that the chiral symmetry breaking is given by  
\begin{eqnarray}
  \label{eq:chiralanom}
  {\cal A}(t,x)
  &=&-\frac{\mu_+}{32\pi^2}\epsilon^{\mu\nu\rho\sigma}{\rm tr}
  F_{\mu\nu}F_{\rho\sigma}+{\rm O}(a)~.
\end{eqnarray}
Since $\mu_+=1$ for $0<m<2r$, this completely agrees with the chiral 
anomaly in the continuum theory. We thus establish that the canonical 
description of the Ginsparg-Wilson fermion correctly reproduces the 
chiral anomaly in the classical continuum limit.

\section{Summary and Discussion}
\label{sec:sd}

We have investigated canonical formulation of massless Dirac theory on 
the spatial lattice. In the absence of gauge 
couplings the theory possesses an exact chiral symmetry of the 
L\"uscher type while avoiding species doubling. The axial charge operator 
commutes with the hamiltonian and the conserved axial charge of a particle 
depends on its momentum. In the classical continuum limit the momentum 
dependence of the axial charge disappears and the states with opposite 
chirality are decoupled with each other. On the lattice, however, the spectra of 
energy and axial charge are smooth periodic functions of momentum with 
a period $4\pi/a$ and 
the gap between positive and negative axial charges disappears 
at the corner points of the first Brillouin zone. Then transitions 
bewteen states with opposite signs of axial charge may occur by the 
gauge couplings. This is responsible for the axial anomaly as was noted 
in Ref. \cite{chn}. 

In the presence of gauge couplings the chiral transformation depends 
on the gauge fields.  The violation of the chiral symmetry in our canonical 
approach is simply due to the fact that the axial charge operator does not 
commute with the hamiltonian. Our computations show that by taking 
account of the fermion loop effect the Ward-Takahashi identity for 
the broken axial charge conservation correctly reproduce the well-known 
anomalous conservation law in the classical continuum limit.

One might think that the GW fermion with the chiral transformation 
(\ref{eq:lctr}) looses superiority to the Wilson fermion with the naive 
chiral transformation (\ref{eq:nctr}) in the presence of gauge couplings. 
The naive chiral symmetry, however, is violated even in the absence of 
gauge couplings whether one uses the Wilson fermion or the GW fermion. 
In the case of the modified chiral trasnformation it is only 
broken at the gauge couplings. We expect that the breaking 
of the modified chiral symmetry is more controllable than that 
of the naive one. This is the virtue of using the GW fermion. 

The interpretation of the axial anomaly as the spectral flow for the adiabatic 
change of the gauge field could be extended to higher dimensions as in the 
continuum theory \cite{nn,Manohar,Fujiwara-Ohnuki}. 
In four dimension we can consider a uniform external 
magnetic field. If the gauge field is time independent, 
we can still define conserved axial charge. The energy spectrum can be 
parametrized by the momentum along the magnetic field. We expect that 
the dispersion relation similar to the two-dimensional case 
discussed in Sect. \ref{sec:cq} arises per flux quantum for 
sufficiently smooth gauge field \cite{Fujiwara-Ohnuki}. 
Applying uniform electric field parallel 
to the magnetic field would induce axial charge nonconservation proportional to 
the total magnetic flux. 

In the euclidean path integral formulation the chiral anomaly is directly 
related to the index of the lattice Dirac operator and gives a topological invariant 
when summed over the lattice. Topologically nontrivial structure of the 
configuration space of lattice gauge fields emerges by imposing admissibility 
condition. This happens even in two dimensions. In the canonical approach the axial 
charge changes continuously. The configuration space of the lattice gauge field is 
divided into sectors by applying the condition that the system approaches, up to 
gauge transformations, to a free field asymptotically. 

It is natural to ask whether the formalism can be applied to chiral gauge theories 
like the standard model. To achieve this it is necessary to define chiral 
fermions. In the euclidean path integral approach fermion field and the 
conjugate field are independent variables and chiral fermions can be
defined by using different chiral projection operators for fermion field 
and the conjugate field. In the canonical approach this does not work 
even for the free theory since the conjugate field is related to the fermion field 
by the Dirac conjugate. This difficulty could be circumvented 
by carefully choosing the fermion contents so that the would-be gauge 
anomalies be canceled. 

Finally the issue of quantizing the gauge field lies beyond the scope of the present 
paper. It is interesting to pursue the understanding of nonperturbative aspects 
of QCD such as the $\theta$-vacuum in the present approach.

\vskip .3cm
We thank Yoshio Kikukawa and Hiroshi Suzuki for valuable  discussions. 
This work is supported in part by the Grant-in-Aid for Scientific Research 
from the Ministry of Education, Culture, Sports, Science and Technology 
(No. 13640258, No. 13135203). 

\appendix

\section{Notation}
\label{sec:not} 
\setcounter{equation}{0}

The metric is assumed to be $\eta^{\mu\nu}={\rm diag}(1,-1,\cdots,-1)$. 
The $\gamma$-matrices satisfy 
\begin{eqnarray}
  \label{eq:gm}
  &&\{\gamma^\mu,\gamma^\nu\}=2\eta^{\mu\nu}~,
  \quad \gamma^0{}^\dagger=\gamma^0~, \quad \gamma^k{}^\dagger=-\gamma^k~, 
  \nonumber \\
  && \gamma_{d+1}=i^{\frac{d}{2}-1}\gamma^0\gamma^1\cdots\gamma^{d-1}~, 
  \qquad \gamma_{d+1}^2=1~.
\end{eqnarray}
In $(1+1)$-dimensions we employ 
\begin{eqnarray}
  \label{eq:gmi2d}
  \gamma^0=\sigma^1=\left(\begin{matrix}0&1\\1&0\end{matrix}\right)~,
  \qquad
  \gamma^1=-i\sigma^2=\left(\begin{matrix}0&-1\\1&0\end{matrix}\right)~,
  \qquad
  \gamma_3=\gamma^0\gamma^1=\sigma^3~.
\end{eqnarray}

Using the forward and backward difference operators $\partial_k$ and 
$\partial_k^\ast$ given by 
\begin{eqnarray}
  \label{eq:fbd}
  \partial_k\psi(t,x)=\frac{1}{a}(\psi(t,x+a\hat k)-\psi(t,x))~, \qquad
  \partial^\ast_k\psi(t,x)=\frac{1}{a}(\psi(t,x)-\psi(t,x-a\hat k))~,
\end{eqnarray}
we define the symmetric and antisymmetric difference operators by
\begin{eqnarray}
  \label{eq:sasdo}
  \partial_k^S=\frac{1}{2}(\partial_k+\partial_k^\ast)~, \qquad
  \partial_k^A=\frac{1}{2}(\partial_k-\partial_k^\ast)~, 
\end{eqnarray}
where $\hat k$ stands for the unit vector along the $k$-th spatial 
coordinate axis. Note that $\partial_k^S\sim{\rm O}(a^0)$ and 
$\partial_k^A\sim{\rm O}(a)$. 

Covariant difference operators are defined by
\begin{eqnarray}
  \label{eq:cdos}
  &&\nabla_k\psi(t,x)=\frac{1}{a}(U_k(t,x)\psi(t,x+a\hat k)-\psi(t,x))~, \nonumber \\
  &&\nabla_k^\ast\psi(t,x)=\frac{1}{a}(\psi(t,x)
  -U_k(t,x-a\hat k)^\dagger\psi(t,x-a\hat k))~. 
\end{eqnarray}
The symmetric and antisymmetric covariant differences are defined by 
\begin{eqnarray}
  \label{eq:csasdo}
  \nabla_k^S=\frac{1}{2}(\nabla_k+\nabla_k^\ast)~, \qquad
  \nabla_k^A=\frac{1}{2}(\nabla_k-\nabla_k^\ast)~. 
\end{eqnarray}

\section{Momentum Integrals}
\label{sec:mi}
\setcounter{equation}{0}

The momentum integrals similar to (\ref{eq:2dic2}) and (\ref{eq:A0}) 
already appeared in the evaluations of chiral anomalies in full 
euclidean lattice theories. They can be carried out by considering a 
continuous map from $p^k\in T^{d-1}$ to 
$z_\mu\in S^{d-1}$ as in Ref. \cite{fns}, where the coordinates on $S^{d-1}$ 
are defined by 
\begin{eqnarray}
  \label{eq:z}
  z^k=-n_k=\rho\sin p^k~, \qquad
  z^d=-n_d~. \qquad (k=1,\cdots,d-1)
\end{eqnarray}
The volume form $d^{d-1}z$ of the $(d-1)$-disk $D^{d-1}$ 
defined by $(z^1)^2+\cdots+(z^{d-1})^2\leq1$ is related 
to the volume form $d^{d-1}p$ of the momentum space by 
\begin{eqnarray}
  \label{eq:vf}
  d^{d-1}z=d^{d-1}pJ(p)~,
\end{eqnarray}
where $J(p)$ is the Jacobian
\begin{eqnarray}
  \label{eq:jac}
  J(p)
  =\frac{\partial(z^1,\cdots,z^{d-1})}{%
    \partial(p^1,\cdots,p^{d-1})}
  =\Bigl(\prod_pm_p\Bigr)n_d\Biggl(
  n_d-r\sum_{p}\frac{n_p^2}{m_p}\Biggr)
\end{eqnarray}
Here we are interested in the integral
\begin{eqnarray}
  \label{eq:I}
  I=I_++I_-~, \qquad
  I_\pm=\int_{-\pi}^\pi d^{d-1}p
  \theta(\mp n_d)f(n_d)\Bigl(\prod_pm_p\Bigr)\Biggl(
  n_d-r\sum_{p}\frac{n_p^2}{m_p}\Biggr)~,
\end{eqnarray}
where $\theta$ is the unit step function. Let $D_+$ ($D_-$) 
be the domain in the momentum space that is mapped into 
the upper (lower) hemisphere of $S^{d-1}$ with $z^d>0$ 
($z^d<0$). Then the insertion of $\theta(\mp n_d)$ 
effectively restricts the momentum region to $D_\pm$. 
Since $n_d=\mp\sqrt{1-r^2}$ with $r=\sqrt{(z^1)^2+\cdots
+(z^{d-1})^2}$ on $D_\pm$, we obtain 
\begin{eqnarray}
  \label{eq:I+}
  I_\pm&=&\mp\mu_\pm
  \int_{D^{d-1}}d^{d-1}z \frac{f(\mp\sqrt{1-r^2})}{\sqrt{1-r^2}}
  \nonumber \\
  &=&\mp\mu_\pm\frac{2\pi^{\frac{d-1}{2}}}{\Gamma((d-1)/2)}
  \int_0^1dr\frac{r^{d-2}f(\mp\sqrt{1-r^2})}{\sqrt{1-r^2}}~,
\end{eqnarray}
where $\mu_\pm$ is defined by 
\begin{eqnarray}
  \label{eq:dm}
  \mu_\pm=\sum_{\{p|z^d(p)=\pm1\}}{\rm sgn}J(p)~. 
\end{eqnarray}
If there is no point on $D_+$ ($D_-$) that is mapped 
to the north pole with $z^d=1$ (the south pole with $z^d=-1$), then 
$\mu_+=0$ ($\mu_-=0$). 

The computation of $\mu_\pm$ is similar to the  counting of the 
doubler modes in Ref. \cite{fns}. We consider the points $(p_\pi^1,\cdots,p_\pi^{d-1})\in 
T^{d-1}$ with $p_\pi^k$ being either $0$ or $\pi$. They are mapped 
to the north pole ($z^d=1$) or the south pole ($z^d=-1$) of $S^{d-1}$  
by (\ref{eq:z}).  For the parameters $r$ and $m$ satisfying 
$2n<m/r<2(n+1)$ ($n=0,\cdots,d-1$),  the points $(p_\pi^1,
\cdots,p_\pi^{d-1})$ where at most $n$ entries are $\pi$ are mapped to 
the north pole, otherwise to the south pole. We thus find 
\begin{eqnarray}
  \label{eq:m+}
  \mu_+=\sum_{s=0}^n(-1)^s{d-1 \choose s}~, 
  \qquad
  \mu_-=\sum_{s=n+1}^{d-1}(-1)^s{d-1 \choose s}~, 
\end{eqnarray}
We see that $\mu_\pm$ satisfy 
\begin{eqnarray}
  \label{eq:mpm2d}
  \mu_++\mu_-=0~. 
\end{eqnarray}
If we assume $\displaystyle{{d-1\choose s}}=0$ for $s<0$ or 
$s>d-1$, (\ref{eq:m+}) becomes valid for any integer $n$. 
It yields $\mu_\pm=0$ for $m<0$ or $m>2d$. 

In two dimensions we find $\mu_\pm=\pm1$ for $0<m/r<2$ and 
$\mu_\pm=0$ for $m<0$ or $m/r>2$. 
Carrying out the remaining radial integral for $\displaystyle{f=\sqrt{
\frac{1+n_2}{2}}}$ and noting (\ref{eq:I}), we obtain 
\begin{eqnarray}
  \label{eq:2di}
  \int_{-\pi}^\pi dp
  \sqrt{\frac{1+n_2}{2}}m_1\Bigl(n_2-r\frac{n_1^2}{m_1}
    \Bigr)=
    \begin{cases}
      -4 &\hbox{for}~~ 0<m/r<2 \cr
      0 &\hbox{for} ~~ m<0,~ m/r>2 \cr
    \end{cases}~. 
\end{eqnarray}

To find the anomaly coefficients appearing in (\ref{eq:A0}) 
we need only two types of momentum integrals $I_1$ and $I_2$ 
defined by 
\begin{eqnarray}
  \label{eq:I1}
  I_1&=&\int_{-\pi}^\pi d^3p
  \frac{1}{\sqrt{2(1+n_4)}}\Biggl(\prod_pm_p\Biggr)\Biggl(
  n_4-r\sum_p\frac{n_p^2}{m_p}\Biggr) \\
  \label{eq:I2}
  I_2&=&\int_{-\pi}^\pi d^3p
  \frac{n_4}{\sqrt{2(1+n_4)}}\Biggl(\prod_pm_p\Biggr)\Biggl(
  n_4-r\sum_p\frac{n_p^2}{m_p}\Biggr)
\end{eqnarray}
In four dimensions $\mu_\pm$ are given by
\begin{eqnarray}
  \label{eq:mpm4d}
  \mu_+=-\mu_-=\begin{cases}
    0 & \hbox{for}~~m<0,~~m/r>6 \cr
    1 & \hbox{for}~~0<m/r<2,~~4<m/r<6 \cr
    -2 & \hbox{for}~~2<m/r<4 \cr 
  \end{cases}
\end{eqnarray}
Using (\ref{eq:I}) and (\ref{eq:I+}) and carrying out the 
radial integrations, we find 
\begin{eqnarray}
  \label{eq:I12}
  I_1=-\frac{16\pi}{3}\mu_+~, \qquad 
  I_2=\frac{16\pi}{15}\mu_+~. 
\end{eqnarray}
\eject

\end{document}